# Application of quasi-steady state photoconductance technique to lifetime measurements on c-Ge substrates

I. Martín, A. Alcañiz, A. Jiménez, G. López, C. del Cañizo, A. Datas

*Abstract*— Similar to other high quality crystalline absorbers, an accurate knowledge of surface passivation of crystalline Germanium (c-Ge) substrates is crucial for a straightforward improvement of photovoltaic device performance. For crystalline silicon devices, this information is typically obtained by quasi-steady state photoconductance (QSS-PC) technique using Sinton WCT-120 tool. In this work, we explore the conditions to adapt this measurement technique to c-Ge substrates. Based on PC-1D simulations, we deduce that a minimum effective lifetime is needed corresponding to an effective diffusion length equal to the substrate thickness. Apart from this, an accurate estimation of the total photogeneration inside the c-Ge sample is also mandatory. This condition implies that the light intensity that impinges onto the sample must be measured with a c-Ge sensor, although the integrated c-Si sensor can be used for high flash intensities. Additionally, the optical factor used to evaluate sample reflectance must be also known, which is determined by measuring robust effective lifetime values under photoconductance decay conditions. Finally, knowledge about carrier mobility in c-Ge is also necessary to translate the measured photoconductance to the corresponding excess carrier density values. Lifetime measurements of passivated c-Ge substrates done by QSS-PC technique are validated by comparing them with the ones obtained by microwave photoconductance technique.

## I. INTRODUCTION

IN the last decades, photovoltaic devices based on crystalline germanium (c-Ge) substrates have been developed for different applications like bottom cells for multijunction III-V space solar cells [1-2], terrestrial applications in concentrated-PV systems [3] and thermoPhotovoltaic (TPV) devices [4]. Similar to other photovoltaic devices based on good quality crystalline materials, surface passivation for c-Ge wafers is crucial for improving c-Ge solar cell efficiencies. Typically, surface recombination in c-Ge is determined by measuring effective lifetime of samples that are symmetrically covered by the passivating films using microwave detected photoconductance decay (μW-PCD) technique [5-7]. In this technique, the sample is illuminated by short laser pulses and photoconductance is monitored through changes in the microwave reflectance of the sample. After every laser pulse, the photoconductance decay is recorded and adjusted by a monoexponential fit resulting in a unique effective lifetime value. This technique shows an important drawback: the dependence of lifetime on excess carrier density cannot be easily measured due to the uncertainty in the illumination intensity per unit of area, which prevents the access to useful information for interface characterization, i.e. the so-called lifetime spectroscopy [8]. Moreover, despite commercial tools are available, they are much expensive and, thus, less common than the Sinton WCT-120 tool, which is based on quasi-steady state photoconductance (QSS-PC) and designed for measuring c-Si substrates. In this technique, a flash lamp illuminates the sample and photoconductance is measured by inductive coupling while light intensity is measured by a calibrated silicon sensor. The knowledge of these two magnitudes permits a straightforward way to measure the evolution of lifetime on excess carrier density. Thus, an adaptation of this technique to c-Ge lifetime measurements would be desirable, as it was already identified in reference [9], where a first approach to this problem was reported. In that work, the authors proposed the modification of carrier mobilities and determined a calibration factor for the photogeneration measured by the silicon sensor by simulations. However, the effect of the non-uniform photogeneration was not addressed which is one of the key points in our work, as it is explained in the following sections.

In this paper, we study the validity of measuring c-Ge substrates using the WCT-120 tool which is based on QSS-PC technique. A thorough analysis of the different aspects involved in the measurement is carried out, identifying the requirements and modifications needed to obtain reliable lifetime values.

Manuscript received January 20th, 2020. This work was partially funded by Ministerio de Ciencia, Innovación y Universidades from Spanish government under projects TEC2017-82305-R and ENE2017-86683-R, and Comunidad de Madrid under project Madrid-PV2 (S2018/EMT-4308). A. Jiménez acknowledges "Programa de Ayudas a la Investigación en Energía y Medio Ambiente 2019" from Fundación Iberdrola.

Isidro Martín, Albal Alcañiz and Gema López are with the Departament d'Enginyeria Electrònica, Universitat Politècnica de Catalunya, 08034 Barcelona, Spain. (e-mail: isidro.martin@upc.edu, albaalcanizmoya@gmail.com, gema.lopez@upc.edu). Alba Jiménez, Carlos del Cañizo and Alejandro Datas are with Instituto de Energía Solar, Universidad Politécnica de Madrid, 28040 Madrid, Spain. (alba.jimenez@ies.upm.es, a.datas@ies.upm.es, carlos.canizo@ies.upm.es).

## II. THE QUASI-STEADY STATE PHOTOCONDUCTANCE TECHNIQUE AND ITS LINK TO SURFACE RECOMBINATION PROPERTIES

Quasi-steady state photoconductance (QSS-PC) technique was proposed by Cuevas and Sinton [10]. In this technique, the light source is a flash lamp whose time decay is much slower than the effective lifetime to measure and, thus, a quasi-steady state is reached for every instant of time. Later, Nagel et al. [11] extended this technique for any time dependence of the flash lamp. Using this general approach, we can calculate the effective lifetime ($\tau_{eff}$) applying the following expression [11]:

$$\tau_{eff}(\Delta n_{av}) = \frac{\Delta n_{av}(t)}{G_{av}(t) - \frac{d\Delta n_{av}(t)}{dt}} \quad (1)$$

where $\Delta n_{av}$ is the average excess minority carrier density and $G_{av}$ is the average photogeneration. These magnitudes are calculated integrating them along the wafer thickness ($w$):

$$G_{av}(t) = \frac{1}{w}\int_0^w G(x,t)dx \quad (2)$$

$$\Delta n_{av}(t) = \frac{1}{w}\int_0^w \Delta n(x,t)dx \quad (3)$$

where $\Delta n(x,t)$ is the excess carrier density and $G(x,t)$ the photogeneration rate. In the WCT-120 tool $\tau_{eff}(\Delta n_{av})$ curves are calculated using equation (1), where $\Delta n_{av}(t)$ and $G_{av}(t)$ are obtained from photoconductance of the wafer ($\Delta\sigma$) and the light intensity of the flash lamp ($I_{ph}$), respectively. The former is measured through a calibrated response of a coil whose magnetic field is coupled to sample conductivity, while the latter is obtained by a calibrated c-Si solar cell located close to the sample under test and integrated in the tool. After flashing the light, both magnitudes are recorded as a function of time. From photoconductance, $\Delta n_{av}(t)$ is calculated using the following expression, in which we have assumed that every photon generates only one electron-hole pair:

$$\Delta n_{av}(t) = \frac{\Delta\sigma(t)}{q(\mu_n + \mu_p)w} \quad (4)$$

where $\mu_n$ ($\mu_p$) is the electron(hole) mobility and $q$ the fundamental charge. On the other hand, the average photogeneration per unit of volume ($G_{av}(t)$) is obtained from the measured intensity ($I_{ph}(t)$) by:

$$G_{av}(t) = \frac{I_{ph}(t) \cdot f_{opt}}{w} \quad (5)$$

where $f_{opt}$ is an optical factor that indicates the fraction of the light measured at the calibrated cell that goes into the sample.

In order to link $\tau_{eff}(\Delta n_{av})$ measured curves to the recombination mechanisms in the sample ($U(x,t)$), we need to integrate all recombination rates inside the wafer:

$$\int_0^w U(x,t)dx = w\frac{\Delta n_{av}(t)}{\tau_{eff}} = \Delta n(0,t) \cdot S_{front} + \int_0^w \frac{\Delta n(x,t)}{\tau_b(\Delta n)}dx + \Delta n(w,t) \cdot S_{rear} \quad (6)$$

Then:

$$\frac{1}{\tau_{eff}} = \frac{\Delta n(0,t)}{\Delta n_{av}(t)} \cdot \frac{S_{front}}{w} + \frac{1}{w\Delta n_{av}(t)}\int_0^w \frac{\Delta n(x,t)}{\tau_b(\Delta n)}dx + \frac{\Delta n(w,t)}{\Delta n_{av}(t)} \cdot \frac{S_{rear}}{w} \quad (7)$$

where $\tau_b$ is the bulk lifetime and $S_{front}$ ($S_{rear}$) is the front (rear) surface recombination velocity. In order to evaluate more easily surface passivation, both wafer surfaces typically are identically processed leading to $S = S_{front} = S_{rear}$. Additionally, this equation can be simplified for symmetrical $\Delta n(x,t)$ profiles, i.e. $\Delta n(0, t) = \Delta n(w, t)$, resulting in identical impact of both surfaces on $\tau_{eff}$. Finally, for well passivated surfaces, $\Delta n(x, t)$ can be approximated as uniform along the wafer [12]. Under these circumstances $\Delta n(x, t) \approx \Delta n_{av}(t)$ and equation (7) results in:

$$\frac{1}{\tau_{eff}(\Delta n_{av})} = \frac{1}{\tau_b(\Delta n_{av})} + \frac{2S(\Delta n_{av})}{w} \quad (8)$$

This equation is widely used when surface passivation is evaluated and using it for c-Ge wafers would be desirable. In principle, an uniform photogeneration along the wafer thickness is needed to maintain a symmetrical $\Delta n(x,t)$ profile and apply equation (8). The flash lamp provided with the tool (QFlash X5dR) is a photographic flash that, despite most of the energy is located at the 900-950 nm band, a significant light intensity is emitted in the visible part of the spectrum. These relatively short wavelength photons are absorbed in the first microns of the c-Si sample altering the symmetrical $\Delta n(x,t)$ profile. To prevent this effect, an optical filter that attenuates wavelengths shorter than 650 nm is also included with the tool. Consequently, the assumption of a constant generation is reasonably fulfilled for c-Si wafers due to their low absorption coefficients at such IR wavelengths.

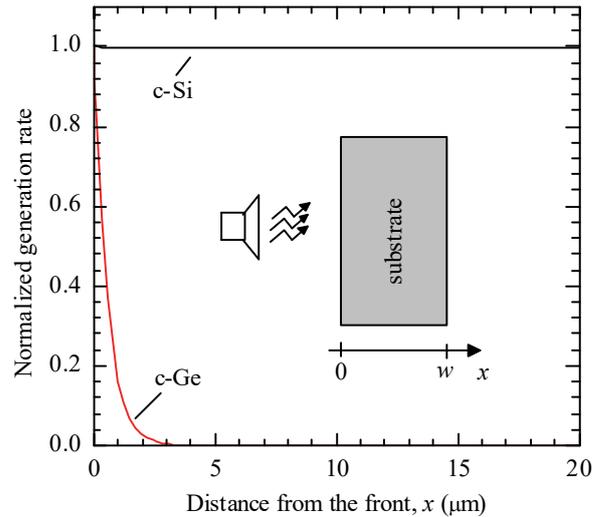

Fig. 1. Generation rate normalized to its value at the surface simulated with PC-1D for a c-Si and c-Ge wafer illuminated by the SintonWCT-120 flash including the IR optical filter.

On the contrary, these conditions are far from been satisfied when a c-Ge wafer is measured with the same lamp: due to its higher absorption coefficients in a broader spectrum, photogeneration of IR light is also concentrated in the first microns. In order to graphically show the different response of the c-Si and c-Ge wafers to the light coming from the flash lamp, in figure 1 we show the generation rate along the first 20 μm of the wafer calculated with PC-1D [13] normalized to its value at the surface. In this calculation, we use the spectrum reported in ref [14], where the effect of the IR pass filter is included, and the default optical properties for c-Ge and c-Si provided with the software. As it can be seen, the generation rate is almost constant for c-Si wafers while it dramatically decreases for c-Ge vanishing in the first 3 μm. The impact of this distortion on the validity of simplified equation (8) for c-Ge substrates measured with the Sinton WCT-120 tool is one of the key points to assess the validity of QSS-PC technique on c-Ge substrates. This analysis together with other relevant factors are addressed in the following section.

### III. REQUIREMENTS FOR c-GE LIFETIME MEASUREMENTS USING WCT-120

#### A. Impact of non-uniform photogeneration

*1) Effect of bulk lifetime*

In order to evaluate the conditions where equation (8) is valid for c-Ge measurements, we reproduce the QSS-PC measurements in PC-1D. As it has been mentioned above, we use the spectrum reported in ref. [14] as primary illumination whose light intensity is defined as an exponential decay with decay time of 40 μs which is the shortest flash duration that can be selected in the flash lamp configuration. As substrates we define a n-type 170 μm- thick c-Ge wafer with a resistivity of 1 Ω·cm, corresponding to $N_D = 1.63 \cdot 10^{15}$ cm$^{-3}$. Mimicking the Sinton WCT-120 tool procedure, from PC-1D we obtain the values of $\Delta\sigma(t)$ and $f_{opt} \cdot I_{ph}(t)$ (the effect of $f_{opt}$ uncertainty is explained below) as a function of time. These values are processed with equations (4) and (5) to get $\Delta n_{av}(t)$ and $G_{av}(t)$ and finally equation (1) is applied to obtain $\tau_{eff}(\Delta n_{av})$ curves, referred to as "simulated experiment" curves in the next figures.

Firstly, we analyse the effect of the bulk lifetime on the QSS-PC measurements cancelling any surface recombination ($S_{front} = S_{rear} = 0$). In figure 2, we show the simulated carrier profile normalized to $\Delta n_{av}$ 50 μs after starting the flash with $\tau_b$ ranging from 10 ns to 100 μs. As it can be seen, for very low $\tau_b$ values the carrier profile is strongly asymmetrical, with very high values close to the surface where the light is entering into the wafer. Notice that carriers are photogenerated at the first 0-3 μm (see figure 1) and due to the short diffusion length, they recombine very close to the place where they were generated. However, higher $\tau_b$ values result in more symmetrical carrier profile since carriers can travel further deep in the wafer. It is obvious that diffusion length ($L_D$) is playing a key role in this dependence. This magnitude is defined as [15]:

$$L_D = \sqrt{D_{min} \cdot \tau_b} \qquad (9)$$

where $D_{min}$ is the diffusion constant of minority carriers that is related to carrier mobility following Einstein's relations [15]:

$$D_{min} = \mu_{min} \frac{k_B \cdot T}{q} \qquad (10)$$

where $\mu_{min}$ is the mobility of minority carriers, $k_B$ is Boltzman's constant and $T$ is the temperature. Taking into account the minority carrier mobility used ($\mu_p = 2445$ cm$^2$/V·s) and the temperature of 300 K, a $\tau_b$ value of 4.57 μs results in $L_D = w$. In other words, this is the minimum $\tau_b$ value to consider that a significant quantity of photogenerated carriers are able to reach the rear surface. In fact, the higher $\tau_b$ the more symmetrical $\Delta n(x)$ profile. As it can be seen in the inset of figure 2, reasonably flat $\Delta n(x)$ profiles are obtained for $\tau_b$ higher than 10 μs, which agrees well with the calculated minimum $L_D$ needed.

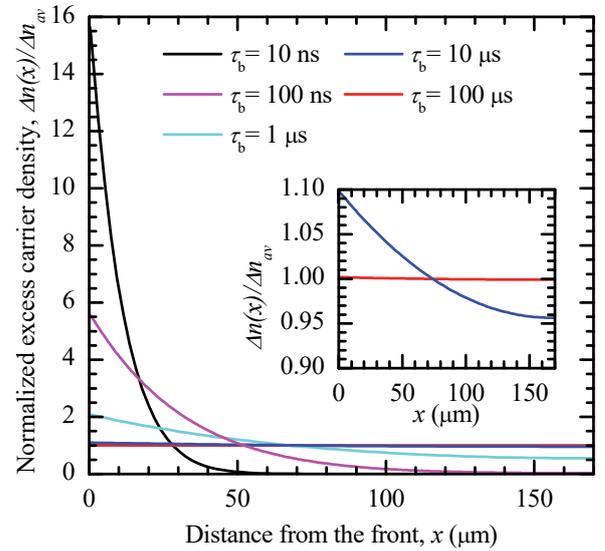

Fig. 2. Normalized excess minority carrier profiles along a n-type 1 Ωcm c-Ge wafer for different bulk lifetimes, 50 μs after starting the flash; in the inset, a zoom of the $\Delta n(x)/\Delta n_{av}$ curves for $\tau_b = 10$ and 100 μs is shown.

*2) Effect of surface recombination velocity*

Now, we focus on the requirements for surface recombination velocity on both surfaces. In this case, for bulk recombination we consider only intrinsic mechanisms through intrinsic lifetime ($\tau_{int}$), i.e. Auger and radiative recombination processes, as modelled in PC-1D [13]. In figure 3, we show the simulated $\tau_{eff}(\Delta n_{av})$ curves (solid lines) for $S_0$ values ranging from $10^6$ to $10^1$ cm/s, where we have defined surface recombination properties for a unique surface state at the intrinsic energy level with equal fundamental recombination velocities for electrons and holes $S_0 = S_{n0} = S_{p0}$. Additionally, we theoretically calculate the $\tau_{eff}(\Delta n_{av})$ curves (dashed lines) using Shockley-Read-Hall equations [16-17] for the calculation of the surface recombination velocity and the corresponding $\tau_{int}$ for bulk recombination to finally combine them through the simplified equation (8).

As it can be seen, simulated experiment and theoretical curves agree well for $S_0$ lower than $10^4$ cm/s corresponding to $\tau_{eff}$ higher than ~1 μs indicating that equation (8) is valid beyond a certain minimum effective lifetime. In order to get a deeper insight, figure 4 shows the minority carrier profile 50 µs after starting the flash illumination. Due to the asymmetrical photogeneration, higher carrier densities are obtained at the front surface. For very high $S_0$ values, apart from the asymmetrical profile close to the front surface, the time that carriers must spend to diffuse to the rear surface is also distorting the $\tau_{eff}$ measurement [12], leading to a mismatch when simplified equation (8) is applied. However, for $S_0$ lower than $10^4$ cm/s reasonably flat carrier profiles are observed leading to a good agreement between simulation and theory (in agreement to what is shown in figure 3).

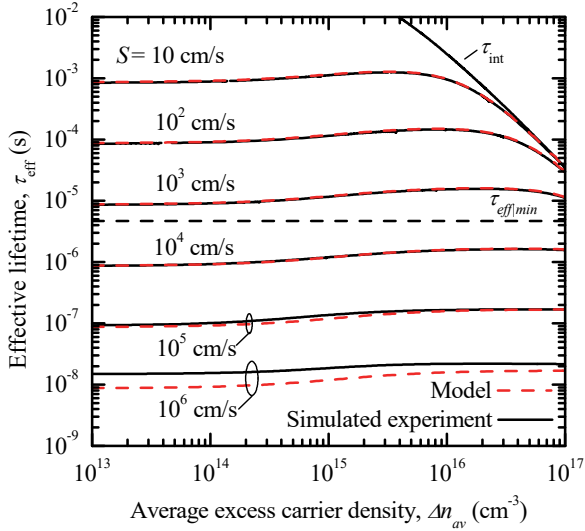

Fig. 3. Comparison of simulated experiment (solid line) and theoretical (dash line) $\tau_{eff}$ vs. $\Delta n_{av}$ curves for effective surface recombination velocities ranging from 10 to $10^6$ cm/s. The corresponding $\tau_{eff|min}$ is also shown. Good agreement is obtained for $\tau_{eff} > 1$ µs, which can be related to a minimum value of $L_{eff}$.

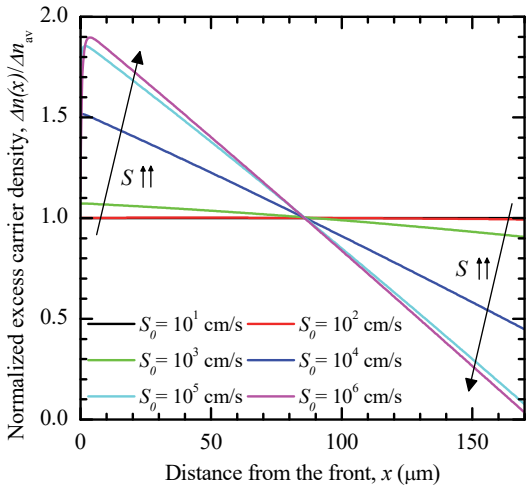

Fig. 4. Normalized excess carrier density profile 50 µs after starting the flash for effective surface recombination velocities ranging from 10 to $10^6$ cm/s. Close to flat profiles are obtained for $S_0 < 10^4$ cm/s.

This result can be also linked to a minimum diffusion length needed. In this case, we should use the effective diffusion length $L_{eff}$ defined as [15]:

$$L_{eff} = \sqrt{D_{min} \cdot \tau_{eff}} \quad (11)$$

Thus, we can claim that the value of $L_{eff}= w$ can be used as a threshold beyond which simplified equation (8) can be accurately applied for illumination sources that lead to a non-uniform photogeneration profile in WCT-120. Notice that this condition is an extension of the one related to the bulk recombination explained above and, thus, it is already included. The minimum valid effective lifetime value is:

$$\tau_{eff}\big|_{min} = \frac{w^2}{D_{min}} \quad (12)$$

The corresponding $\tau_{eff|min}$ value is also plotted in figure 3 confirming that $\tau_{eff}$ values beyond that value can be accurately related to $S$ and $\tau_b$ using equation (8). In fact, as it can be seen in figure 3, theoretical and simulated experiment curves converge each other as $\tau_{eff}$ approaches $\tau_{eff|min}$ leading to acceptable results even for slightly lower $\tau_{eff}$ values.

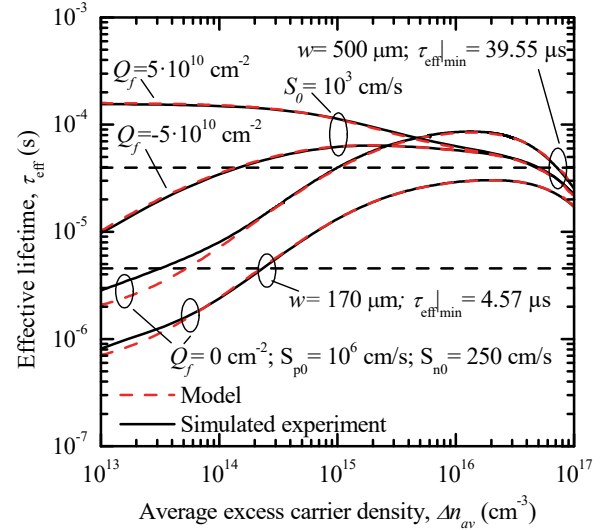

Fig. 5. Comparison of simulated experiment (solid lines) and theoretical (dash lines) $\tau_{eff}$ vs. $\Delta n_{av}$ curves for different wafer thicknesses and surface configurations. As it can be seen, the analytical model can be applied with negligible error for $\tau_{eff}$ values higher than $\tau_{eff|min}$ corresponding to an effective diffusion length longer than the wafer thickness.

In order to evaluate to applicability of this condition, figure 5 shows simulation and theoretical curves of a sample for different wafer thicknesses and/or surface configurations. In particular, we define asymmetrical fundamental surface recombination velocities ($S_{n0}= 250$ cm/s and $S_{p0}=10^6$ cm/s) or positive and negative fixed charge density ($|Q_f|= 5 \cdot 10^{10}$ cm$^{-2}$), that lead to an inverted and accumulated surface, respectively. For the theoretical calculation of the $\tau_{eff}(\Delta n)$ curves with $Q_f \neq 0$ cm$^{-2}$, we use Girisch model [18]. The value of $\tau_{eff|min}$ is also plotted for the two simulated thicknesses, namely 170 and 500 µm. As it can be seen, excellent agreement between theoretical curves applying equation (8) and simulated experiment ones are obtained for $\tau_{eff}$ values higher than the corresponding $\tau_{eff|min}$ value. Although good agreement is also observed for $\tau_{eff}$ values slightly lower than $\tau_{eff|min}$, beyond this point theoretical and

simulated curves tend to diverge with significant differences when $\tau_{eff}$ is much lower than $\tau_{eff|min}$.

In summary, all the results presented in this section demonstrate that $\tau_{eff}$ values higher than $\tau_{eff|min}$ permit the application of the widely-used relation shown in equation (8) despite of the non-uniform photogeneration profile. Moreover, apart from the c-Ge used in these simulations, this condition can be applied to many different interesting photovoltaic crystalline materials like III-V semiconductors.

*3) Transient evolution of excess carrier density profile*

Up to now, we have considered that the excess carrier density profile is approximately uniform for $\tau_{eff}$ longer than $\tau_{eff|min}$. However, just after starting the flash, the photogeneration and, thus, the excess carrier density is concentrated at the front surface to be subsequently dispersed along the wafer. As a consequence, there must be some time at the beginning of the flash illumination where the homogeneous excess carrier density profile is not fulfilled and $\tau_{eff}$ cannot be accurately measured. In figure 6, we show the simulated excess carrier density normalized to its average value at different times after starting the flash, namely: 1, 2.5, 5, 10 and 25 μs. In the inset, the simulated curves of illumination intensity and photoconductance signals are also shown (lines) with the symbols indicating the corresponding times where the excess carrier density profile is shown. All these simulations are done with only intrinsic recombination mechanisms in the bulk and $S_0 = 10^3$ cm/s. This $S_0$ value results in $\tau_{eff}$ value close to $\tau_{eff|min}$ (see figure 3), which can be considered as a worst case.

As it is expected, after 1 μs $\Delta n(x)$ is very asymmetrical since carriers do not have had time to flow to the rear surface. In addition, as it can be seen in the inset of figure 6 photoconductance signal is increasing indicating that recombination at the surfaces is not able to equilibrate photogeneration, which means that more excess carrier density is needed. This increasing trend is maintained until photoconductance signal shows a maximum where excess carrier density is enough to make recombination compensate photogeneration. Notice that beyond the time when photoconductance signal reaches its maximum, carrier profile is dominated by recombination characteristics of the sample that are defined symmetrical, in contrast to the asymmetrical photogeneration rate. In figure 6, this condition occurs at 25 μs after starting the flash where we can see an almost flat $\Delta n(x)$. Simulations show that for samples with $\tau_{eff} > \tau_{eff|min}$, the $\Delta n(x)$ profile is close to homogeneous at the time when photoconductance is maximum. Notice that the longer $\tau_{eff}$ values, the longer the time to reach this maximum since higher excess carrier densities are needed and the longer available time for carriers to be distributed along the wafer. As a consequence, for correct lifetime measurements we must use signal values after photocoductance has reached its maximum value. In fact, this procedure is also common when c-Si samples are measured.

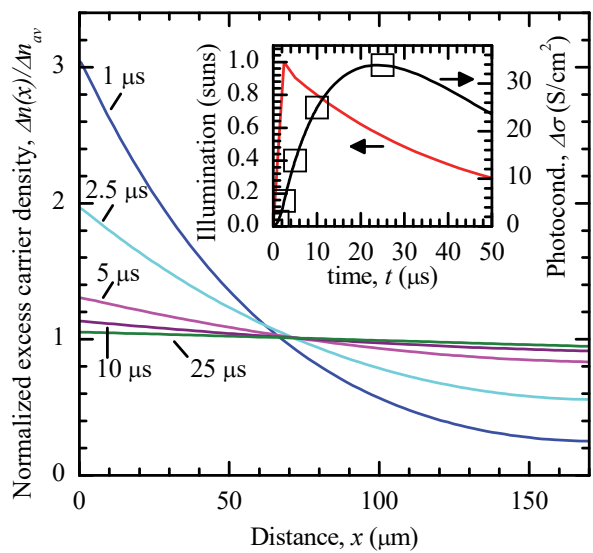

Fig. 6. Evolution of excess carrier density profile when photoconductance signal is increasing. An almost homogeneous carrier profile is obtained when photoconductance reaches its maximum. In the inset, photoconductance (black line) and illumination (red line) signals are shown with symbols indicating the times where the carrier profiles are simulated.

Apart from this requirement, simulations extended beyond that time show a weak dependence of carrier profile on flash duration. In figure 7, we show carrier profiles after 50 μs for short (decay time of 40 μs) and long (decay time of 2 ms) flashes with the same recombination properties than in the previous figure ($\tau_b = \tau_{int}$ and $S_0 = 10^3$ cm/s). We can see that more symmetrical profiles are obtained for short flashes. The origin of this difference is attributed to the fact that light intensity after 50 μs of starting the flash is much lower for short flashes than for long flashes. With short flashes, instantaneous photogeneration is not so dominant since most part of the carriers comes from previous illumination intensities, i.e. they have not recombined yet. This effect is stronger for lower $S_0$, i.e. longer effective lifetimes, reducing the difference between both flash durations (not shown in the graph). The limit case is when a pulse of light is used, coming for example from a LED or laser light source, where a perfectly symmetrical profile is obtained, as it is shown in figure 7. In fact, in this case the symmetrical profile is achieved very quickly as it is shown in the inset of figure 7 where we also show carrier profiles for this type of light excitation at 1, 2 and 3 μs after abruptly turning the light off. As it can be seen, after only 3 μs perfectly symmetrical profile is obtained due to the absence of light. As a consequence, simplified equation (8) can be applied to effective lifetime measurements carried out by μW-PCD technique, as it was already identified in reference [5].

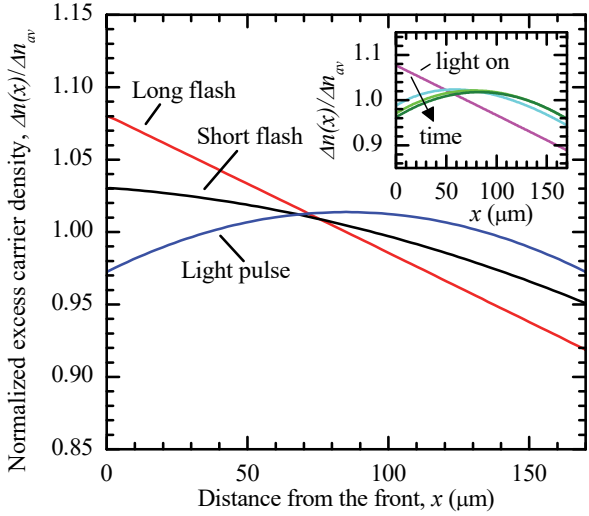

Fig. 7. Normalized excess carrier density profile after 50 μs starting the flash or after finishing light pulse. As it can be seen profile is more symmetrical for shorter flashes being the light pulse the limiting case. In the inset, we show carrier profiles when the illumination is on and after 1, 2 and 3 μs of turning off a light pulse. This time is enough to get a symmetrical carrier profile that validates the application of equation (8) for μW-PCD measurements.

### B. Light sensor

The WCT-120 tool has an integrated c-Si solar cell that is used to have a signal proportional to the light intensity that impinges sample surface creating the excess carrier density. When a c-Si sample is measured, this sensor responds to the same wavelength range than the sample under test leading to reliable measurements. However, if the integrated sensor is used when c-Ge samples are measured, photons in the IR can generate electron-hole pairs in the c-Ge sample while they do not in the c-Si sensor. This spectrum mismatch can lead to a systematic error in the measurement of the light intensity that is relevant for the c-Ge sample.

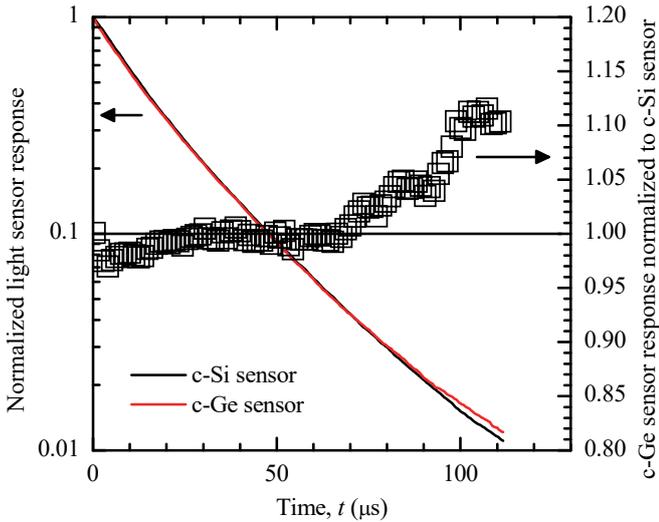

Fig. 8. c-Ge and c-Si sensor responses normalized to the maximum value and the ratio between them. Spectrum components with lower energies than c-Si bandgap are dominant in the last part of the flash.

In order to evaluate this error, we simultaneously measure flash light intensity with the IR pass filter with the integrated c-Si solar cell and an external sensor of c-Ge. In figure 8, we show both signals normalized by their corresponding maximum value obtained at $t = 0$. As it can be seen, very similar response is obtained with both sensors except for the last part of the flash where the c-Ge sensor gives a slightly higher signal. As long as flash intensity decays, IR components become more dominant creating a mismatch between sensor responses with higher signal for the c-Ge one. The magnitude of this deviation can be evaluated by dividing both signals, also plotted in figure 8, where factors of about 1.1-1.15 are obtained for the last part of the flash.

Based on these results, we can see that the integrated c-Si sensor can be used limiting the acquired data to the first 75 μs after its maximum. However, although c-Si sensor can still partially work, a c-Ge sensor is desirable in order to fully use all the measured data. In addition, taking advantage that for very high flash intensities both sensor responses are similar, we can use this part of the signal to determine the calibration factor of the c-Ge sensor that converts the measured volts to light intensity.

### C. Carrier mobilities

In order to accurately calculate $\Delta n_{av}$ from the measured photoconductance, a detailed knowledge about carrier mobilities is needed when applying equation (4). For c-Si, dependence of mobilities on doping densities [19] and carrier-to-carrier scattering [20-21] are included in the software provided by the manufacturer with the WCT-120 tool. For c-Ge substrates, accurate measurements of carrier mobilities as a function of doping densities and temperature can be found in the literature [22-25]. However, as far as we know, experimental data of the effect of carrier-to-carrier scattering on mobility is not available for c-Ge (only a theoretical approach can be found in reference [26]). As a first approach, we use the model included in PC-1D where only the dependence on doping densities is taken into account using the following equation, with values for $\mu_{max}$, $\mu_{min}$, $\alpha$ and $N_{ref}$ shown in table I taken from the mentioned references:

$$\mu(N_{dop}) = \frac{\mu_{max} - \mu_{min}}{1 + \left(\frac{N_{dop}}{N_{ref}}\right)^\alpha} - \mu_{min} \quad (13)$$

TABLE I
PARAMETER VALUE FOR MOBILITY MODEL

|  | $\mu_{max}$ (cm$^2$ V$^{-1}$·s$^{-1}$) | $\mu_{min}$ (cm$^2$ V$^{-1}$·s$^{-1}$) | $N_{ref}$ (cm$^{-3}$) | $\alpha$ |
|---|---|---|---|---|
| Electrons | 3895 | 641 | 6.13·10$^{16}$ | 1.04 |
| Holes | 2505 | 175 | 9.33·10$^{16}$ | 0.9 |

### D. Optical factor

Once the light intensity that reaches the sample is measured, we need to know how much light is able to get into it and photogenerate carriers to calculate the corresponding $G_{av}(t)$ using equation (5). In that expression, the optical properties of the sample are summarized in an optical factor ($f_{opt}$). To determine its value, we use a solid state light source that permits an accurate variation of the light intensity. This fine tuning of the excitation light has demonstrated to be effective for $f_{opt}$ determination [27]. In our case, the sample is illuminated by an 850 nm LED array with a pulsed light that leads to Photoconductance Decay (PCD) conditions. In other words, once light is switched off a symmetrical carrier profile is obtained after few microseconds (see inset of figure 6) and the effective lifetime can be calculated using equation (1) with $G_{av}$=0:

$$\tau_{eff}(\Delta n_{av}) = -\frac{\Delta n_{av}(t)}{\frac{d\Delta n_{av}(t)}{dt}} \qquad (14)$$

These lifetime values are very robust since we only need to have a linear response of the photoconductance sensor, which is fulfilled by the one included in WCT-120. Effective lifetime values determined with this technique can be used to calibrate $f_{opt}$ that impacts on the $G_{av}$ value when the flash lamp is used. In figure 9 we show $\tau_{eff}(\Delta n)$ curves measured with the flash lamp (symbols) and with the LED array (line). As it can be seen, the data measured with the flash lamp is impacted by the different $f_{opt}$ values with a good overlapping with LED array measurement for $f_{opt} = 0.93$. We estimate that this method allows the determination of the optical factor with an accuracy of ±0.05.

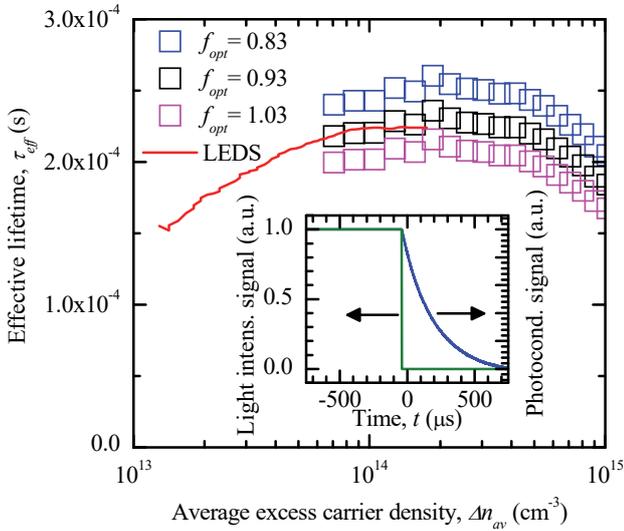

Fig. 9. $\tau_{eff}$ vs. $\Delta n_{av}$ measurement of c-Ge sample with flash lamp (symbols) and LED array (solid line). The latter permits to adjust the optical factor ($f_{opt}$); in the inset, we show the signals obtained for light intensity and photoconductance using pulsed light for the LED array.

Despite this technique can be applied to any sample, it must be mentioned that the LED array could not be necessary if $\tau_{eff}$ values are much longer than flash duration. In that case, PCD conditions will be fulfilled after finishing the flash light permitting the adjustment of $f_{opt}$ [28].

IV. VALIDATION OF THE LIFETIME MEASUREMENTS

Based on the conclusions reached in the previous section, we measured $\tau_{eff}(\Delta n)$ curves of a <100> 170 μm-thick 1 Ωcm n-type c-Ge wafer symmetrically passivated by PECVD silicon carbide films (a detailed description of the results regarding the optimization of the passivation properties of these films will be published elsewhere). The RF bridge was automatically calibrated before lifetime measurement using air conductivity. The only modifications to conventional c-Si lifetime measurements consisted of introduction of c-Ge mobilities and, although it is not mandatory, the use of a c-Ge sensor for light intensity measurement. The obtained data is validated by comparing the obtained results with the measurement carried out with Semilab WT-2000 which uses the μW-PCD technique. In this tool, a head with an excitation laser and a sensor to measure sample photoconductance through microwave coupling scans the sample leading to a lifetime mapping. For every scanned position, only one lifetime value is obtained by fitting the exponential decay of photoconductance losing the information about the level of excess carrier density.

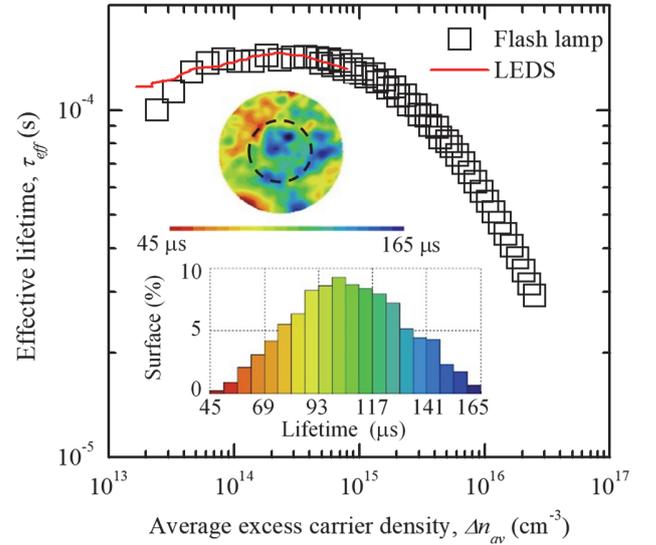

Fig. 10. Comparison of $\tau_{eff}$ vs. $\Delta n_{av}$ curves of passivated c-Ge sample measured with flash lamp and LED array with Sinton WCT-120 together with the lifetime mapping measured by Semilab WT-2000 and the corresponding lifetime distribution. Values between both techniques agree well in the range of $10^{13}$-$10^{15}$ taking into account that the magnetic sensor of WTC-120 measures a 2 cm diameter circular area in the centre of the wafer.

In figure 10, we show the $\tau_{eff}(\Delta n)$ curves measured with the Sinton WCT-120 using the flash lamp and the LED array. As it was explained in the previous section, overlapping both curves allow us to determine the optical factor $f_{opt}$. Together with these curves, we show the mapping of the same sample measured with Semilab WT-2000 and the statistical distribution of the lifetime values. Although average lifetime value is 106 μs taking into account the whole sample, sensitivity of the magnetic coil is located at a circumference with a radius of about 0.9-1 cm [29]. This high sensitivity region is indicated in the lifetime mapping and the measured $\tau_{eff}$ values in this region agree well with the lifetime values measured with Sinton WCT-120 which shows a plateau of 140 µs for $\Delta n_{av}$ = $10^{14}$-$10^{15}$ cm$^{-3}$ which is the estimated $\Delta n_{av}$ range used in the lifetime mapping [30]. This good agreement demonstrates the validity of $\tau_{eff}(\Delta n_{av})$ measurements using QSS-PC technique on c-Ge samples.

## V. CONCLUSION

In this work, we have demonstrated that measurements of $\tau_{eff}(\Delta n)$ based on QSSPC technique using Sinton WCT-120 are reliable if the following requirements are fulfilled. Firstly, effective diffusion length must be longer than the thickness of the c-Ge substrate so that excess carrier density profile is flat enough to involve both surfaces into the effective lifetime value. And secondly, an accurate knowledge of the photogeneration inside the sample is needed. This requirement implies that the sensor that measures light intensity must respond to the same light spectrum of the sample. Thus, a c-Ge sensor is desirable, although the integrated c-Si sensor can also partially work for high flash intensities. In addition, the optical factor, which is related to the reflectance properties of the sample, is also crucial. In this work, this factor has been determined by means of a pulsed light using a LED array that results in robust $\tau_{eff}$ values that can be used to calibrate the QSS-PC measurement. Apart from these requirements, to calculate excess carrier density from photoconductance both electron and hole mobilities in c-Ge must be applied. Precise values of those parameters would be desirable in order to improve measurement accuracy. The validity of the measurement technique is demonstrated by comparing results of QSS-PC measurements of a passivated c-Ge sample to lifetime data obtained by µW-PCD with good matching between them.